\documentclass[12pt]{article}

\usepackage{graphicx}
\usepackage{subfigure}

\input tcilatex
\QQQ{Language}{
American English
}

\begin{document}

\title{Facilitated alkali ion transfer at the water 1,2-dichloroethane interphase:
Ab-initio calculations concerning alkaline metal cation -
1,10-phenanthroline complexes.}
\author{Cristi\'{a}n G. S\'{a}nchez, Ezequiel P. M. Leiva \\
Unidad de Matem\'{a}tica y Fisica,\\
Facultad de Ciencias Qu\'{\i}micas,\\
Universidad Nacional de C\'{o}rdoba\\
Agencia Postal 4, CC 61, 5000 C\'{o}rdoba, Argentina \and Sergio Dassie, Ana
M. Baruzzi \\
Departamento de Fisicoqu\'{\i}mica,\\
Facultad de Ciencias quimicas,\\
Universidad Nacional de Cordoba\\
Agencia Postal 4, CC 61, 5000 C\'{o}rdoba, Argentina}
\maketitle

\begin{abstract}
A series of calculations on the energetics of complexation of alkaline
metals with 1,10-phenanthroline are presented. It is an experimental fact
that the ordering of the free energy of transfer across the water -
1,2-dichloroethane interphase is governed by the charge / size ratio of the
diferent cations; the larger cations showing the lower free energy of
transfer. This ordering of the free energies of transfer is reverted in the
presence of 1,10-phenanthroline in the organic phase. We have devised a
thermodynamic cycle for the transfer process and by means of ab-initio
calculations have drawn the conclusion that in the presence of phen the free
energy of transfer is governed by the stability of the PHEN/M $^{+}$complex,
which explains the observed tendency from a theoretical point of view.
\end{abstract}

\section{Introduction}

Cation transfer across liquid/liquid interphases assisted by complexing
agents has been the subject of intensive research in the last years \cite
{d1, d2, d3, d4, d5}. One main reason is the application of this phenomenon
in cation extraction processes \cite{d6, d7, d8} and in the development of
new ion selective electrodes. In this sense, alkaline and alkaline-earth
cations transfer is mainly relevant in the field of clinical chemistry. Many
natural and synthetic ligands have been reported \cite{d9, d10} and are
continuously appearing in an attempt to improve selectivity, increase
stability constants of the complex and solubility of the above mentioned
ions in the organic phase. Among these compounds, 1,10-phenanthroline is a
very well known ligand, especially for transition metals, which are highly
soluble in water and have very low selectivity. Alkaline and alkaline-earth
cations complexes with 1,10-phenanthroline are also known. Cyclic
voltammetry applied to four-electrode systems has been used \cite{d11} to
analyze the transfer mechanism, nature and stoichiometry of the complexes.
The experimental data indicates that in the case of Li$^{+}$ the transfer
would occur according to the sequence:

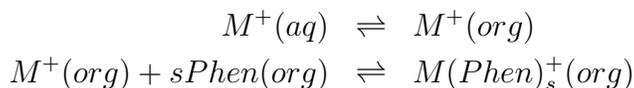
\begin{eqnarray*}
M^{+}(aq) &\rightleftharpoons &M^{+}(org) \\
M^{+}(org)+sPhen(org) &\rightleftharpoons &M(Phen)_s^{+}(org)
\end{eqnarray*}
where the indices ``aq'' and ``org'' indicate aqueous and organic phases
respectively and $s$ denotes the stoichiometry of the metal-phenanthroline
complex. In the case where no 1,10-phenanthroline (Phen from now on) is
added to the organic phase, the potential of transfer and accordingly the
free energy of transfer is lower for larger alkaline cations according to
the smaller charge/size ratio for these ions. In the presence of Phen in the
organic phase this order in the free energy of transfer is reverted
indicating a strong interaction of the ions with smaller radii with Phen.

Synthesis and characterization of 1,10-phenanthroline derivatives have been
performed in the last years by different groups \cite{d9, d10, d11}, the
main aim of this research being the development of a Li$^{+}$ selective
electrode for monitoring Li$^{+}$ activity in biological systems. In this
respect, calculation of the interaction energy between ligand and cation
would be helpful for the prediction and design of new ionophores.

It is the purpose of the present work to provide a qualitative explanation
for some of the above mentioned experimental results concerning the transfer
of different alkaline ions through the water/1,2-dichloroethane interphase
facilitated by 1,10-phenanthroline.

\section{The Model}

\begin{figure}
  \centering
  \includegraphics[scale= .35,angle=90]{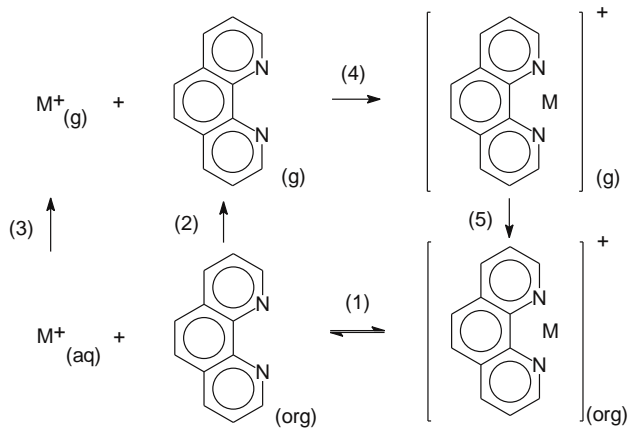}
  \caption{Thermodynamic cycle used to analyze facilitated transfer
of alkaline ions assisted by 1,10-phenanthroline.}
\end{figure}

In order to explain the experimental results from a thermodynamic point of
view we devised a thermodynamic cycle for the overall transfer process as
shown in Figure 1. The global transfer process, here denoted with (1) is
divided into a number of elementary steps (2-5) for which we are going to
analyze the relative contributions to (1). In step (2) we take
phenanthroline, which is dissolved in the organic solvent to the gas phase.
We shall denote this contribution with -$\Delta G^{dis}$ . In step (3) we
take an alkaline ion, which is dissolved in the aqueous phase to the vacuum.
This step involves a free energy change which is equal to \textit{minus} the
solvation free energy of the ion in water $-\Delta G_{solv}^{M^{+}}$. In
step (4) we combine the alkaline ion with phenanthroline in the gas phase to
yield a compound that we shall denote as $PhenM^{+}$.We label the related
free energy change with $\Delta G_f^{PhenM^{+}}$ . In step (5) we bring $%
PhenM^{+}$ into the organic phase, with an associated free energy change $%
\Delta G_{solv}^{PhenM^{+}}$.

In order to consider relative differences between the transfer free energies
of the different alkaline ions, we can leave aside the contribution of step
(2) (which amounts 68.9 KJ/mol) to the total free energy change, since it is
the same for all the ions. If the contribution $-\Delta G_{solv}^{M^{+}}$ of
step (3) were the dominant one, the overall process would be favored for the
bigger ions, as most solvation models state, this desolvation energy is
inversely proportional to the radius of the ion. Turning to step (5), the
resolvation of the complex should involve only a minor contribution to the
total free energy, since the charge/size ratio of this species should be
almost the same for all the ions. Thus, step (4) must be responsible for the
observed trend in the free energy of transfer. In order to check out the
validity of this conclusion, we performed a series of ab-initio calculations
of the structure of alkaline metal - 1,10-phenanthroline cationic complexes.
We analyzed the differences of energy of the complexation process between
the different ions in terms of the nature of the chemical bond that is
formed.

\section{Ab-initio calculations}

The subject of the present study are the complexes of 1,10-phenanthroline
with lithium, sodium and potassium cations (Li$^{+}$, Na$^{+}$, K$^{+}$).
The first step of the study was to determine the structure of the different
complexes. Geometry optimizations were carried out for 1,10-phenanthroline,
LiPhen$^{+}$,NaPhen$^{+}$ and KPhen$^{+}$ using STO-3G \cite{sto1}\cite{sto2}%
\cite{sto3} and MIDI \cite{MIDI} basis sets at the RHF level, the energies
of the single ions Li$^{+}$, Na$^{+}$and K$^{+}$ were also calculated. The
results for these energy calculations are shown in Table 1. MP2 corrections $%
\Delta E_c$ accounting for the correlation contribution to the total energy
were also calculated and added to the RHF/MIDI results, as given in the
third column of this table. These corrections include core orbitals and were
obtained using the geometries at the energy minimum of the RHF/MIDI results.
All calculations were performed at $C_{2v}$ symmetry.

Table 2 shows the results for the energy change associated with complex
formation $\Delta E_{cf}(PhenM^{+})$ at the different levels of theory as
calculated from the results from Table 1. In all cases we find the stability
of the $PhenM^{+}$ complex to increase in the order: $%
PhenK^{+}<PhenNa^{+}<PhenLi^{+}$, in agreement with the expectations
formulated in the previous section.

Taking into account that the basis sets that were used for the calculation
are small, we must consider that the basis set superposition error (BSSE)
associated with the RHF calculation could deliver an important contribution
to the energy change of complex formation. The energy differences related to 
$\Delta E_{BSSE}$, defined below,\ are shown in Table 3 for the RHF
calculations with STO-3G and MIDI basis sets, as calculated according to the
method of Boys and Bernardi \cite{counterpoise}. Within this method, the
corrected energy of complex formation $\Delta E_{cf}^{BSSE-corr}$ is
calculated from:

\[
\Delta
E_{cf}^{BSSE-corr}=E_{cf}(PhenM^{+})-E_{Phen}^{M^{+}-gh}-E_{M^{+}}^{Phen-gh} 
\]
where $E_{PhenM^{+}}$ is the calculated energy for the complex, $%
E_{Phen}^{M^{+}-gh}$ denotes the energy of Phen when the calculations are
performed including the basis functions of M$^{+}$ as a ''ghost'' molecule
and $E_{M^{+}}^{Phen-gh}$ is the energy of M$^{+}$ with the Phen molecule as
a ghost. The BSSE is calculated as follows: 
\[
\Delta E_{BSSE}=\Delta E_{cf}^{BSSE-corr}-\Delta E_{cf}^{uncorr} 
\]

Since the BSSE error arises from the mathematical fact that the basis sets
are not complete, the inclusion of the ghost molecules always produces a
lowering of the energy. $\Delta E_{BSSE}$ is then a positive quantity which
will be lower the larger the basis set used for the calculation. Thus, $%
\Delta E_{cf}^{uncorr}$ includes some energy that is not related to the
chemical reaction itself but with the better description of the wave
function of one monomer augmented by the inclusion of the basis set of the
other. We find that in the present case $\Delta E_{BSSE}$ amounts less than
39\% and 12\%\ of the binding energy $\Delta E_{cf}(PhenM^{+})$ for the
RHF/STO-3G and RHF/MIDI cases respectively.

Another factor to include in the binding energies of $PhenM^{+}$ are the
corrections for the change in zero point energy (ZPE) between Phen and PhenM$%
^{+}$. These ZPE values were obtained from a normal mode analysis at the
equilibrium geometry at RHF/STO-3G and RHF/MIDI levels of theory, further
details about vibrational analysis are given below in section 3. The values
for the ZPE correction are given in Table 4. Since upon complex formation
new weak bonds are formed, the change in ZPE is always positive and small
compared to the total $\Delta E_{cf}(PhenM^{+})$, amounting less than 1.7 \%
and 2.4\%\ of the binding energy for the RHF/STO-3G and RHF/MIDI cases
respectively.

In Table 5 we show the results for the energy change of complex formation
for the different complexes with all corrections included.

Using the theoretical values for the vibrational frequencies, optimized
geometries and energy changes calculated ab-initio, we made estimations for
the changes in the thermodynamic functions for alkaline metal cation
complexation in the gas phase with 1,10-phenanthroline. In Table 6 we show
the results for the changes in internal energy, enthalpy, entropy and free
energy calculated using normal mode harmonic oscillator, rigid rotor and
ideal gas partition functions for the different complexes, metals and Phen.
We find that the entropy change is small, arising fundamentally from the
loss in translational degrees of freedom upon complexation. Thus, the free
energy change of this reaction is energetically dominated at normal
laboratory temperatures.

In spite of the fact that the approximations used here make these values
only a rough estimation of the real ones, we can clearly see a trend in
stability of the different complexes in the order Li \TEXTsymbol{>} Na 
\TEXTsymbol{>} K. According to the thermodynamic cycle shown in section 1,
this would indicate that the free energies of transfer and accordingly the
transfer potential for the different ions across the water
1,2-dichloroethane interphase assisted by 1,10-phenanthroline should follow
the sequence:

\[
\Delta V_K>\Delta V_{Na}>\Delta V_{Li}
\]
which is in fact, as we have previously stated, the order found in the
voltammetric experiment.

\section{Structure and vibrational spectra of the complexes}

\begin{figure}
  \centering
  \subfigure[]{
   \label{fig:subfig:a}
   \includegraphics[width=0.3\textwidth,angle=90]{fig2a.epsi}}
  \hspace{0.5in}
  \subfigure[]{
   \label{fig:subfig:b}
  \includegraphics[width=0.3\textwidth,angle=90]{fig2b.epsi}}
  \subfigure[]{
   \label{fig:subfig:c}
  \includegraphics[width=0.3\textwidth,angle=90]{fig2c.epsi}}
  \hspace{0.5in}  
  \subfigure[]{
   \label{fig:subfig:d}
\includegraphics[width=0.3\textwidth,angle=90]{fig2d.epsi}}
  \caption{RHF/MIDI structure of Phen and PhenM$^{+}$ complexes. All
distances are in \AA .}
\end{figure}

\begin{figure}
  \centering
  \subfigure[1,10-phenanthroline-Li$^+$]{
   \label{fig:subfig:a}
   \includegraphics[width=0.25\textwidth,angle=-90]{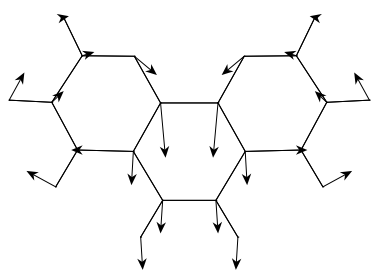}}
  \hspace{0.25in}
  \subfigure[1,10-phenanthroline-Na$^+$]{
   \label{fig:subfig:b}
  \includegraphics[width=0.25\textwidth,angle=-90]{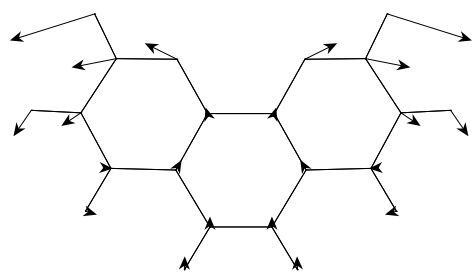}}
  \subfigure[1,10-phenanthroline-K$^+$]{
   \label{fig:subfig:c}
  \includegraphics[width=0.25\textwidth,angle=-90]{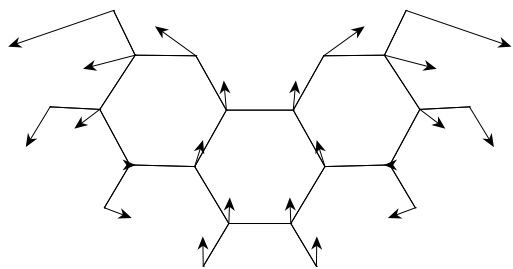}}
  \caption{Perturbation of Phen structure upon complex formation
with the different cations.}
\end{figure}

Figure 2 shows the structure of Phen and Li$^{+}$, Na$^{+}$ and K$^{+}$
complexes as obtained from RHF/MIDI geometry optimizations. The structure of
1,10-phenanthroline is in very good agreement with reported experimental
results based on single crystal X-ray diffraction \cite{phen-rx}. The
structures of complexes PhenLi$^{+}$ and PhenNa$^{+}$ obtained from the
calculations show shorter N-M$^{+}$ bond lengths than those obtained from
X-ray diffraction experiments \cite{rx-li, rx-na1, rx-na2}, this is
reasonable since in the crystal there is always a counterion coordinated
with the metal cation. Comparison of Phen deformation upon complexation is
difficult because in the crystals the Phen molecule is strongly distorted
from C$_{2v}$ symmetry and it is not easy to determine which geometry
changes are produced by complexation and which by crystal packing.

As we may see at a first glance the major structural difference between the
different complexes is the Phen-cation distance, which increases in going
from Li$^{+}$ to K$^{+}$ as shown in Table 7, where we report the N-M$^{+}$
distances $d_{N-M^{+}}$ for the different complexes. It is found that d$%
_{N-M^{+}}$ is smaller for Li$^{+}$ and almost the same for Na$^{+}$ and K$%
^{+}$ complexes.

Figure 3 shows the perturbation of Phen structure produced by the
complexation with the different cations. The arrows indicate the direction
and relative magnitude of the difference in position between corresponding
atoms in the two structures, complexed and free, according to the vector: 
\[
\Delta \mathbf{r}_i=\mathbf{r}_i^{PhenM^{+}}-\mathbf{r}_i^{Phen}
\]
where $\mathbf{r}_i^{PhenM^{+}}$ represents the position vector of atom $i$
in the complex and $\mathbf{r}_i^{Phen}$ the position vector of atom $i$ in
Phen. In order to fix the relative positions between the two molecules for
comparison, the coordinates of the different atoms were set so as to
minimize the sum of the distances between corresponding atoms, 
\[
f(\mathbf{\{\QTR{mathbf}{r}}_i\mathbf{\}})=\dsum_i\left| \mathbf{r}%
_i^{PhenM^{+}}-\mathbf{r}_i^{Phen}\right| 
\]
subject to the constraint of keeping constant the structure of both
molecules.

According to this figure we may think of the Phen molecule as a pair of
forceps that closes around the ion. The presence of Li$^{+}$ between the
forceps originates a closing whereas the presence of Na$^{+}$ or K$^{+}$
originates an opening of the forceps. This observation and the other stated
in the previous paragraph may be briefly stated in the following words:
complexation occurs when the metal accommodates itself in the ''cavity ''
formed by the two nitrogen atoms in Phen. Since Li$^{+}$ is the smallest, it
fits comfortably in this cavity and results in its contraction. On the other
hand, Na$^{+}$ and K$^{+}$ are bigger than the size of the cavity, forcing
Phen to open for a proper accommodation of the incoming ion.

We turn now to the study of the differences in vibrational spectra between
the different complexes involving alkaline ions and Phen. These vibrational
spectra were calculated using normal mode analysis at the equilibrium
geometry at RHF/MIDI level using force constants calculated from ab-initio
analytical second order derivatives of the RHF/MIDI energy. No scaling was
done of vibrational frequencies since no predictive analysis is pretended
but only a qualitative description of the main changes observed. In figure 4
we can see the simulated spectra. In order to clarify the main features we
have superimposed to the usual line representation a graph obtained by
summing gaussian functions centered at each frequency with the corresponding
intensity, all of the same amplitude in wavenumber. This construction does
not attempt to compare with experimental information but to make easier to
detect the main features of the spectra.

\begin{figure}
  \centering
  \includegraphics[scale= .35,angle=-90]{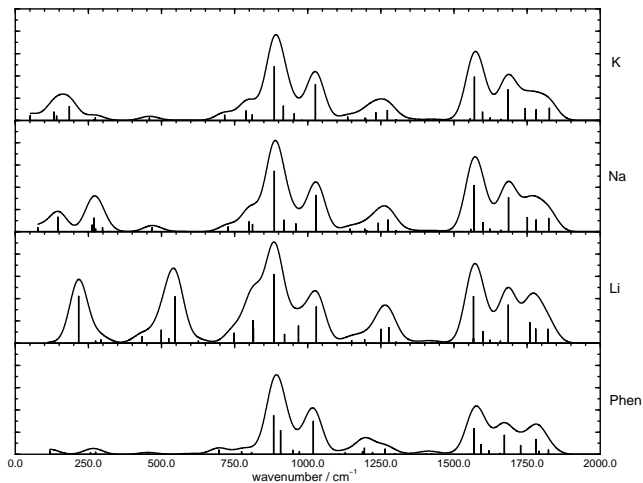}
  \caption{Theoretical non-scaled RHF/MIDI vibrational spectrum of
Phen and PhenM$^{+}$ complexes. The absorbances are expressed in Debye$^2$
amu$^{-2}$\AA $^{-2}$.}
\end{figure}

The fundamental differences between the spectra for the different complexes
appear mainly in the low frequency bands. Two distinctive bands at 250 cm$%
^{-1}$ and 600 cm$^{-1}$ appear in the case of PhenLi$^{+},$ corresponding
to Li out of plane bending and Li-Phen stretching respectively.
Corresponding bands of PhenNa$^{+}$ and PhenK$^{+}$ are displaced towards
considerably lower frequencies indicating the stronger bonding of Li$^{+}$
to Phen with respect to Na$^{+}$ and K$^{+}$ yielding higher force constants
for Li$^{+}$-Phen vibrations.

\section{Nature of the metal-ligand bond}

\begin{figure}
  \centering
  \subfigure[1,10-phenanthroline-Li$^+$]{
   \label{fig:subfig:a}
   \includegraphics[width=0.35\textwidth,angle=-90]{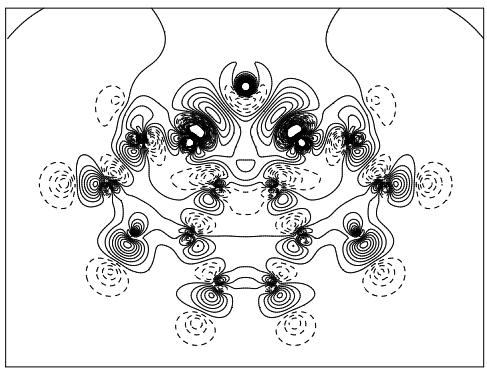}}
  \hspace{0.25in}
  \subfigure[1,10-phenanthroline-Na$^+$]{
   \label{fig:subfig:b}
  \includegraphics[width=0.35\textwidth,angle=-90]{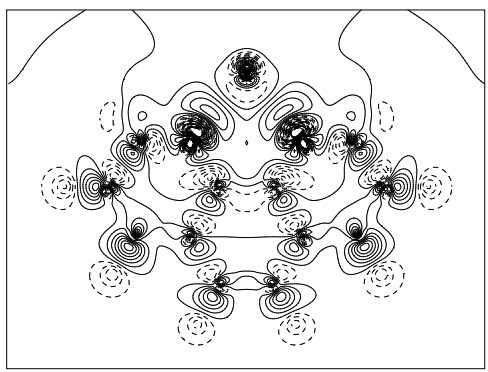}}
  \subfigure[1,10-phenanthroline-K$^+$]{
   \label{fig:subfig:c}
  \includegraphics[width=0.35\textwidth,angle=-90]{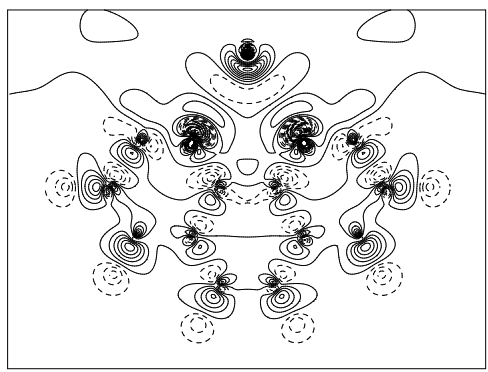}}
  \caption{Electron density difference map upon complex formation
for the different complexes. Continuous contour lines indicate positive $%
\Delta \rho $ and broken contours indicate negative $\Delta \rho $.}
\end{figure}

Electronic density difference maps over the $\sigma _v(xz)$ plane are shown
in Figure 5. These maps represent the difference of electronic density
between the complex and the monomers at the equilibrium geometry of the
complex: 
\[
\Delta \rho (x,z)=\rho ^{PhenM^{+}}(x,z)-\rho ^{Phen}(x,z)-\rho ^{M^{+}}(x,z)
\]

It can be appreciated that the electron withdrawal by the metal ion produces
an electron depletion located mainly in the neighborhood of the N lone pair.
The increase of positive charge on the N atoms sparks a redistribution of
the electronic density via inductive effects across the bonding network.
Thus, a concomitant polarization of the C-H bonds occurs in all the
molecule. However, this has only a minor effect on the C-H vibrational
frequencies and bond distances.

\begin{figure}
  \centering
  \setlength{\abovecaptionskip}{40pt}
  \includegraphics[scale= .35,angle=90]{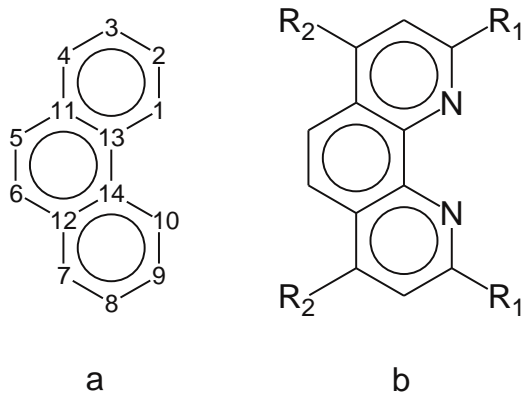}
  \caption{a) numbering scheme used for C atoms, b) Possible
positions for selectivity increasing substituents.}
\end{figure}

Mulliken analysis of the atomic charge on the different cations is shown in
Table 8. The numbering scheme used here for the carbon atoms is shown in
Figure 6a. The electronic charge transfer from Phen to the ion seems to be
rather small, and almost the same for the three species considered here.
According to Mulliken analysis, the carbon atoms which show the largest
positive charge accumulation $\Delta q=q_{\func{complex}ed}-q_{free}$ are
C2, C4 and C13 (and the symmetrical counterparts C9, C7 and C14). In the
most relevant case of Li, the sequence found is the sequence $\Delta q_{C2}$ 
$>\Delta q_{C13}>\Delta q_{C4}$. The C2 atom is directly affected by the
electron withdrawal at the N atom, remaining its only possibility of
recovering electronic charge via inductive effects from C3. On the other
hand, C13 accommodates its charge better since it can take charge from two
bonds instead of only one. The positive charge of C4 is produced by electron
withdrawal through the external and internal bonding network via C11 and C3.

\begin{figure}
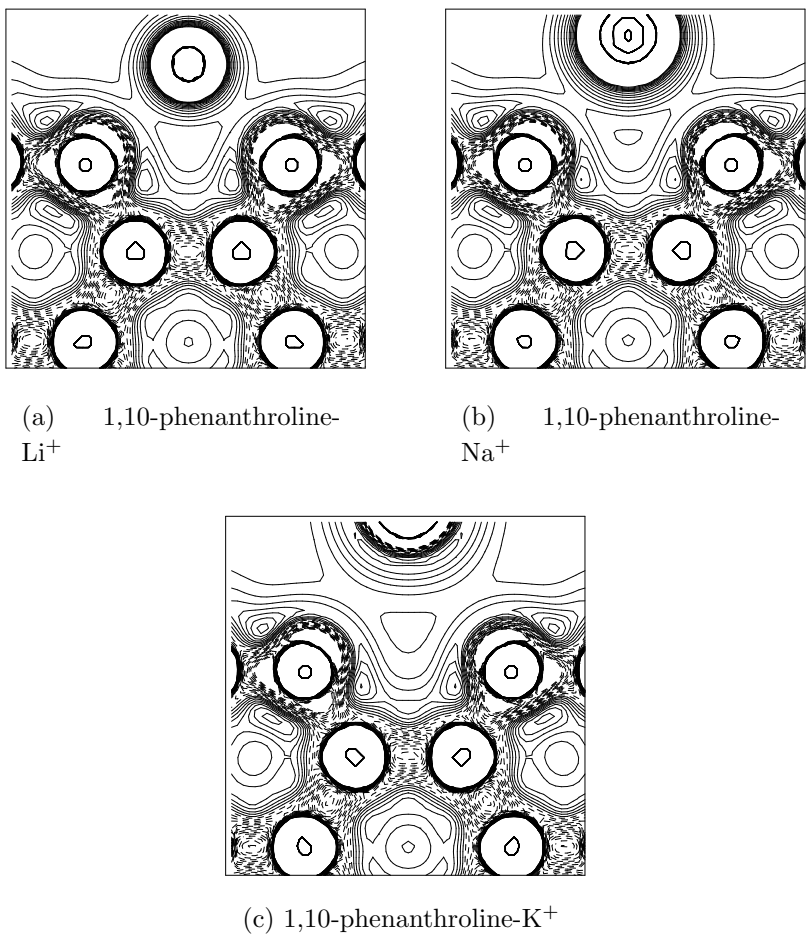

  \centering
  \subfigure[1,10-phenanthroline-Li$^+$]{
   \label{fig:subfig:a}
   \includegraphics[width=0.35\textwidth,angle=-90]{fig7a.epsi}}
  \hspace{0.25in}
  \subfigure[1,10-phenanthroline-Na$^+$]{
   \label{fig:subfig:b}
  \includegraphics[width=0.35\textwidth,angle=-90]{fig7b.epsi}}
  \subfigure[1,10-phenanthroline-K$^+$]{
   \label{fig:subfig:c}
  \includegraphics[width=0.35\textwidth,angle=-90]{fig7c.epsi}}
  \caption{Laplacian of the electronic density for Phen and the
different complexes. Solid contour lines indicate positive $\nabla ^2\rho $
and broken contours indicate negative $\nabla ^2\rho $.}
\end{figure}

As thoroughly analyzed by Bader\cite{Bader}, the laplacian of the electronic
density $\nabla ^2\rho $ can also deliver relevant information concerning
the nature of the chemical bond and chemical reactivity. It can be shown
that if $\nabla ^2\rho <0$ at the point $\mathbf{r}$ , then $\rho (\mathbf{r}%
)$ is greater than the average of its value over an infinitesimal sphere
centered on \textbf{r}, while the opposite is true if $\nabla ^2\rho >0.$
Thus, the laplacian can be used as a criterion for charge concentration or
depletion in the different molecular regions, and it defines in the
interatomic regions the so called valence-shell charge concentration. This
corresponds to the portion of the outer shells of the atoms over which $%
\nabla ^2\rho <0,$ that is, the region where the chemically relevant valence
electronic charge is concentrated. The contourplot of the laplacian of the
electronic density $\nabla ^2\rho $ over $\sigma _v(xz)$ for the different
complexes and Phen are shown in Figure 7 for the region surrounding the N
atoms. The analysis of the plots of the laplacian electronic density for the
different complexes and Phen shows the bonded charge concentration between
the C-N and C-C bonds, as well the non-bonded charge concentration in the
neighborhood of the N atoms. Comparison between the plots obtained for a
free and for a complexed Phen shows that this molecule is not affected by
complex formation even in its finer details and no evidence of a new
chemical bond between Phen and the different cations can be found.
Furthermore no drastic differences are seen in the plots for the different
complexes that cannot be ascribed to the differences in the relative radii
of the different cations. Thus, all previously stated observations suggest
that the Phen - M$^{+}$ bond is mainly of electrostatic nature and that the
orders of stability of the different complexes is produced by the
differences in ionic radii between the different cations.

\section{Conclusions}

We have used a thermodynamic model to analyze the transfer potential of
alkaline ions across water / 1,2-dichloroethane facilitated by
1,10-phenanthroline. Based on ab-initio calculations we have concluded that
this process is dominated by the free energy of complex formation.

The differences in energy and structure of the different complexes can be
explained by the different radii of the cations since the Phen - M$^{+}$
bond is fundamentally of electrostatic (charge - dipole) nature.

According to Mulliken analysis, the carbon atoms that accommodate most of
the positive charge upon complex formation are C2, C13 and C4 (and their
symmetrically equivalent positions). According to this, substitution in C2
and C4 with inductive electron donors R$_2$ and R$_1$ as shown in Figure 6b
would increase the selectivity of Phen towards Li$^{+}$. R$_1$ should be
voluminous so as to prevent the accommodation of the larger cations by
steric hindrance increasing Li$^{+}$ selectivity. The voluminous part of the
substituent should not be located at the C atom vicinal to the ring but
rather in the second C atom so as to produce stronger disturbance for Na and
K which are located farther than Li in the equilibrium structures.
Experimental evidence indicates that 2,9-dibutyl-1,10-phenanthroline shows
excellent selectivity coefficients towards Li$^{+}$ \cite{d13, jap2};
substitution at positions 4 and 7 with phenyl rings increases this
selectivity but not very strongly. This fact supports our conclusions and
leads us to state that it should be possible to predict qualitative
tendencies in selectivities of different ligands performing \emph{\
ab-initio }and probably semiempirical quantum mechanical calculations on the
energetics of complexation. This must be done taking into account that, at
least for the singly charged cations, the complexation step of the cation
plays a dominant role in the energetics of transfer. In the case of doubly
charged cations, the solvation energy increases considerably in magnitude
(it scales with the square of the charge) and may have a more important
contribution to the thermodynamics of the ion transfer. This may explain why
2,9-dibutyl-1,10-phenanthroline is selective towards Li$^{+}$ even in the
presence of Mg$^{2+}$.

\section{Acknowledgments}

All calculations were done in a DIGITAL AlphaStation workstation donated by
Alexander von Humboltd Foundation (Germany). A fellowship (C.S.) from the
Consejo de Investigaciones Cient\'{\i}ficas y T\'{e}cnicas de la Provincia
de C\'{o}rdoba, financial support from the Consejo Nacional de
Investigaciones Cient\'{\i}ficas y T\'{e}cnicas, the Secretar\'{\i}a de
Ciencia y T\'{e}cnica de la Universidad Nacional de C\'{o}rdoba and language
assistance by Pompeya Falc\'{o}n are also gratefully acknowledged.\newpage\ 

\begin{center}
\begin{tabular}{|r|r|r|r|}
\hline
& RHF/STO-3G & RHF/MIDI & MP2/MIDI$^{*}$ \\ \hline
Phen & -560.982204 & -564.572053 & -565.813693 \\ \hline
LiPhen$^{+}$ & -568.346231 & -571.901364 & -573.135921 \\ \hline
NaPhen$^{+}$ & -720.808051 & -725.382428 & -726.620163 \\ \hline
KPhen$^{+}$ & -1154.095224 & -1160.922835 & -1162.045787 \\ \hline
Li$^{+}$ & -7.135447 & -7.185210 & -7.185210 \\ \hline
Na$^{+}$ & -159.657383 & -160.697965 & -160.697965 \\ \hline
K$^{+}$ & -593.009334 & -596.277816 & -596.277816 \\ \hline
\end{tabular}

Table 1: RHF energies of Phen, alkaline cations and MPhen$^{+}$ complexes at
optimized geometries. All values are given in hartrees. $^{*}$at RHF/MIDI
optimized geometry.\vspace{1.0in}

\begin{tabular}{|c|c|c|c|}
\hline
& RHF/STO-3G & RHF/MIDI & MP2/MIDI \\ \hline
$\Delta E_{cf}$ PhenLi$^{+}$ & -0.228578 & -0.144101 & -0.137017 \\ \hline
$\Delta E_{cf}$ PhenNa$^{+}$ & -0.168462 & -0.112410 & -0.108504 \\ \hline
$\Delta E_{cf}$ PhenK$^{+}$ & -0.103685 & -0.072965 & -0.045723 \\ \hline
\end{tabular}

Table 2: Energies of MPhen$^{+}$ complex formation at the different levels
of theory. All values are given in hartrees.\vspace{1.0in}

\begin{tabular}{|c|c|c|}
\hline
& RHF/STO-3G & RHF/MIDI \\ \hline
PhenLi$^{+}$ & 0.058545 & 0.005279 \\ \hline
PhenNa$^{+}$ & 0.058744 & 0.013077 \\ \hline
PhenK$^{+}$ & 0.040848 & 0.005865 \\ \hline
\end{tabular}

Table 3: Basis set superposition errors in the energy of MPhen$^{+}$ complex
formation. All values are given in hartrees.\vspace{1.0in}

\begin{tabular}{|c|c|c|}
\hline
& RHF/STO-3G & RHF/MIDI \\ \hline
LiPhen$^{+}$ & 0.003932 & 0.004489 \\ \hline
NaPhen$^{+}$ & 0.002371 & 0.002738 \\ \hline
KPhen$^{+}$ & 0.001594 & 0.001706 \\ \hline
\end{tabular}

Table 4: Change in zero point energy upon MPhen$^{+}$ complex formation at
the different levels of theory. All values are given in hartrees.%
\vspace{1.0in}

\begin{tabular}{|c|c|c|}
\hline
& RHF/STO-3G & RHF/MIDI \\ \hline
PhenLi$^{+}$ & -0.166101 & -0.134333 \\ \hline
PhenNa$^{+}$ & -0.107347 & -0.096594 \\ \hline
PhenK$^{+}$ & -0.061242 & -0.065393 \\ \hline
\end{tabular}

Table 5: Energy change upon MPhen$^{+}$ complex formation. These values
include BSSE and ZPE corrections. All values are given in hartrees.%
\vspace{1.0in}

\begin{tabular}{|c|c|c|c|c|}
\hline
Ion & $\Delta U/kJ\ mol^{-1}$ & $\Delta H/kJ\ mol^{-1}$ & $\Delta S/J\
mol^{-1}K^{-1}$ & $\Delta G/kJ\ mol^{-1}$ \\ \hline
Li$^{+}$ & -354.38 & -356.86 & -121.43 & -320.66 \\ \hline
Na$^{+}$ & -253.41 & -255.89 & -118.10 & -220.68 \\ \hline
K$^{+}$ & -170.24 & -172.71 & -108.07 & -140.49 \\ \hline
\end{tabular}

Table 6: Theoretical values for the thermodynamic functions of reaction of
an alkaline ion $M^{+}$ with a phenanthroline molecule $Phen$ in the gas
phase to give $M(Phen)_{}^{+}$ at 298K, using RHF/MIDI geometries and
vibrational frequencies.\vspace{1.0in}

\begin{tabular}{|c|c|}
\hline
& N-M$^{+}$ distance \\ \hline
LiPhen$^{+}$ & 1.921 \\ \hline
NaPhen$^{+}$ & 2.218 \\ \hline
KPhen$^{+}$ & 2.264 \\ \hline
\end{tabular}

Table 7: N-M$^{+}$ distances in the different complexes in \AA .%
\vspace{1.0in}

\begin{tabular}{|c|c|}
\hline
& q$_{M^{+}}^{Mull}$ \\ \hline
LiPhen$^{+}$ & 0.93 \\ \hline
NaPhen$^{+}$ & 0.94 \\ \hline
KPhen$^{+}$ & 0.96 \\ \hline
\end{tabular}

Table 8: Cation charge q$_{M^{+}}^{Mull}$according to Mulliken atomic
populations over M in the different complexes.\newpage\ 
\end{center}

\end{document}